\documentclass[lettersize,journal]{IEEEtran}
\usepackage{amsmath,amsfonts,amsthm}

\usepackage{graphicx}
\usepackage{textcomp}
\usepackage{enumerate}
\usepackage{wrapfig}
\usepackage{float}

\usepackage{comment}
\usepackage{comment}
\usepackage{listings,color}
\usepackage{multirow}
\usepackage{balance}
\usepackage{cite}
\usepackage{url}
\usepackage{array,color,colortbl,graphicx,multirow}
\usepackage{subcaption}
\usepackage{caption}
\usepackage{makecell}
\usepackage{url}
\usepackage{nicematrix}
\usepackage{mathtools}
\usepackage{tikz}
\usetikzlibrary{shapes}
\usetikzlibrary{decorations.markings,shadows, shapes}
\usepackage[linesnumbered,ruled,vlined]{algorithm2e}
\SetKwInput{KwData}{Input}
\SetKwInput{KwResult}{Output}
\SetKwComment{Comment}{/* }{ */}
\usepackage{wrapfig}
\usepackage{enumitem}
\usepackage{booktabs}
\usepackage{marvosym}

\usepackage{etoolbox}
\usepackage{tikz}
\usetikzlibrary{patterns, positioning,shapes.misc,matrix, arrows.meta, calc} %
\usepackage[most]{tcolorbox}  
\usepackage{xcolor} 
\usepackage{siunitx}
\usepackage{tabularx}
\usepackage{amssymb}
\usepackage{orcidlink}

\renewcommand{\arraystretch}{0.92}
\usepackage[compact]{titlesec}

\titlespacing*{\section}
{0pt}{0.65ex plus 0.2ex minus 0.2ex}{0.45ex plus 0.1ex}

\titlespacing*{\subsection}
{0pt}{0.55ex plus 0.2ex minus 0.2ex}{0.35ex plus 0.1ex}

\titlespacing*{\subsubsection}
{0pt}{0.45ex plus 0.2ex minus 0.2ex}{0.25ex plus 0.1ex}

\usepackage{etoolbox}

\usepackage[alignedforall]{lpform}
\usepackage{epigraph}

\hypersetup{
    colorlinks=true,
    citecolor=blue,
    linkcolor=blue,
}

\begin{document}

%\title{Gemini Scheduling for Distributed DNN Training}

\title{RingStitch: Demand-Aware Optical Stitching for Fragmented TPU Clusters}

%\author{IEEE Publication Technology,~\IEEEmembership{Staff,~IEEE,}

\author{
    \IEEEauthorblockN{
    Huiru Ao$^1$, 
    Fan Yang$^{1}$, 
    Binglei Wang$^1$,  
    Bo Liu$^1$, 
    Jialong Li$^{1,}$\textsuperscript{\Letter}
    } 
    \\
    \IEEEauthorblockA{\normalsize \textsuperscript{1} Faculty of Computer Science and Artificial Intelligence, Shenzhen University of Advanced Technology\\
    }
    \IEEEauthorblockA{\textsuperscript{\Letter}\text{\texttt{lijialong@suat-sz.edu.cn} 
    }}
}

% \markboth{Transactions on Networking,~Vol.~XX, No.~XX, AUGUST~2025}%

\maketitle

\begin{abstract}
Large-scale AI training clusters increasingly use optical circuit switching (OCS) to reconfigure rack-level interconnects and create elastic accelerator slices. In multi-tenant TPU-style clusters, however, small and medium jobs often leave partial free capacity stranded inside racks. Although the aggregate free capacity may be sufficient for a new job, it cannot be used by local placement or coarse full-rack stitching. This paper presents RingStitch, an OCS-based defragmentation scheduler that turns fragmented rack capacity into schedulable resources. RingStitch follows a local-first policy, stitches compact cross-rack fragments only when needed, and orders the selected racks into a low-cost logical ring. Simulations on a TPU 8t-like SuperPod model show that RingStitch improves schedulability over local and full-rack baselines. 
%Large-scale AI training clusters increasingly use optical circuit switching (OCS) to assemble elastic accelerator slices. In multi-tenant TPU-style systems, however, small and medium jobs leave residual capacity dispersed across partially occupied racks. This capacity remains inaccessible to rack-local placement and coarse full-rack stitching even when sufficient in aggregate. We present RingStitch, an OCS-aware scheduler that recovers these fragments while limiting communication cost. When local placement fails despite sufficient aggregate capacity, RingStitch generates fragment- and topology-aware cross-rack candidates and selects a compact rack set and low-cost ring order.
%上面是一个更短的版本，不过语言不太流畅。
At 95\% load on three fragmentation-heavy workloads, RingStitch improves utilization by up to 5.2 percentage points and reduces demand-weighted blocking ratio by up to 7.8 percentage points relative to Full-Rack Stitching. Its average and P95 waiting-time reductions reach 91.6\% and 94.7\%, respectively, while limiting cross-rack communication cost.
%我仅改动了原版的数据分析部分，但是有点长，先保留。
%注意：修改后图片的位置可能会改变。
\end{abstract}


\section{Introduction}

Large language models and multimodal foundation models increasingly rely on training clusters with thousands of AI accelerator chips. Google TPU systems are representative: TPU v4 introduced optically reconfigurable rack-level interconnects~\cite{jouppi2023tpu}, and TPU 8t scales a SuperPod to 9600 TPU chips~\cite{tpu8t}. At this scale, scheduling depends not only on aggregate idle capacity, but also on whether the available chips can be assembled into an efficient training slice.

Optical circuit switching (OCS) adds a powerful placement primitive: a scheduler can select racks and configure optical links for a job-specific slice. This capability reflects a broader move toward datacenter networks whose connectivity adapts to demand~\cite{wang2010c,farrington2010helios,chen2013osa,farrington2013multiport,mellette2017rotornet,mellette2020expanding,ballani2020sirius}. In TPU-style systems, however, this flexibility can be restricted to coarse allocation units: a job either fits within one rack or receives a slice of fully available racks. This works when job size aligns with rack capacity, but it leaves residual capacity stranded when multi-tenant workloads partially occupy many racks.

We focus on this rack-level fragmentation problem. A cluster may have enough idle chips in aggregate, while no single rack and no set of fully free racks can host the job. The capacity exists, but it is unusable under the allocation granularity. Local First-Fit and Best-Fit are trapped by the single-rack boundary; full-rack stitching ignores partial free capacity; and naive cross-rack aggregation can scatter a job across distant racks and inflate communication cost. The scheduler should therefore recover stranded fragments without turning capacity gain into communication overhead.

Existing ML-cluster schedulers improve time sharing, job completion time, fairness, goodput, heterogeneity support, and network-aware placement~\cite{xiao2018gandiva,gu2019tiresias,mahajan2020themis,qiao2021pollux,narayanan2020heterogeneity,peng2018optimus,zhao2020hived,rajasekaran2024cassini}. These systems optimize how jobs share accelerators over time or where jobs should be placed under conventional cluster abstractions. They do not model OCS as a placement-time resource for stitching partial rack capacity into communication-aware training slices.

This paper presents RingStitch, an OCS-based defragmentation scheduler for multi-tenant TPU-style clusters. RingStitch treats partial free rack capacity as a first-class schedulable resource. It follows a local-first policy and invokes optical stitching only when local placement fails but aggregate residual capacity is sufficient. It then selects compact, fragment-aware rack sets, allocates chips with a fragment-first policy, and orders the selected racks into a low-cost logical ring. The result is a practical balance between resource utilization and communication efficiency.

The main contributions are as follows. First, we formulate rack-level residual fragmentation in OCS-enabled TPU clusters by distinguishing aggregate capacity from policy-feasible capacity. Second, we design RingStitch, which combines bounded candidate generation, fragment-first allocation, and topology-aware ring ordering under a lexicographic objective. Third, we evaluate RingStitch using a TPU 8t-like SuperPod simulator under diverse workloads.
%On fragmentation-heavy workloads, RingStitch improves utilization and reduces the blocking ratio and waiting time relative to Full-Rack Stitching, while limiting cross-rack communication cost through topology-aware candidate selection.
%注意：修改后图片的位置可能会改变。这句话和abstract的最后一句话非常相似，只是没有加入具体数字。

\section{Motivation}
\label{sec:motivation}

\begin{figure*}[t]
    \centering
    \includegraphics[width=\textwidth]{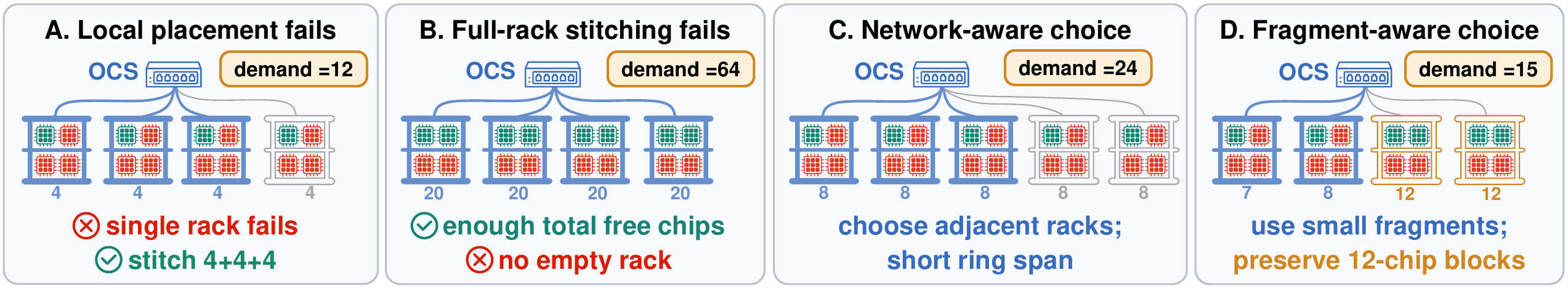}
    \caption{Four motivating placement cases. Values below racks denote residual capacity in chips. Blue racks form the selected cross-rack placement; orange racks in Case D are deliberately preserved.}
    \label{fig:motivation}
\end{figure*}  

Fig.~\ref{fig:motivation} separates placement feasibility from placement quality. In Case A, no rack can accommodate a 12-chip job, although three 4-chip fragments provide exactly the required capacity. Case B shows why full-rack stitching remains insufficient: four partially occupied racks provide 80 free chips in aggregate, enough for a 64-chip job, but none is fully idle. Both cases require stitching residual capacity from partially occupied racks. However, feasibility alone does not determine a good placement. In Case C, each candidate rack has eight free chips, but selecting three topologically nearby racks yields a lower closed-ring cost than using a dispersed set. In Case D, when rack count and ring cost are otherwise comparable, consuming the 7- and 8-chip fragments exactly satisfies the 15-chip demand and preserves both 12-chip residual blocks for later rack-local placement. Together, these cases motivate a local-first scheduler that aggregates residual capacity when rack-local placement fails while controlling rack count, ring cost, and post-placement fragmentation.

\section{System Modeling and Scheduling Semantics}

\begin{table}[t]
    \small
    \centering
    \color{black}
    \arrayrulecolor{black}
    \caption{\textcolor{black}{Summary of main notations and parameters.}}
    \label{tab:table1}
    \begin{tabular}{ll}
    \toprule
    \textbf{Symbol} & \textbf{Meaning} \\
    \midrule
    $N$ & Number of racks in the cluster \\
    $C$ & Number of chips per rack \\
    $f_i$ & Residual free chips in rack $i$ \\
    $d_j$ & Chip demand of job $j$ \\
    $a_j$ & Arrival time of job $j$ \\
    $t_j$ & Execution duration of job $j$ \\
    $S_j$ & Rack set selected for job $j$ \\
    $x_{j,i}$ & Chips allocated from rack $i$ to job $j$ \\
    $c(u,v)$ & Torus distance between racks $u$ and $v$ \\
    \bottomrule
    \end{tabular}
\end{table}

\subsection{Cluster and Rack Abstraction}\label{sec:cluster_model}
Table~\ref{tab:table1} summarizes the main notation used throughout the model. We model a TPU-style OCS cluster as a rack set $\mathcal{I}=\{1,\ldots,N\}$. Each rack contains $C$ chips, for a total capacity of $NC$. Its residual capacity is an integer $f_i\in\{0,\ldots,C\}$. Racks occupy a fixed $M_x\times M_y\times M_z$ three-dimensional torus, where $M_xM_yM_z=N$, and rack $i$ has coordinate $p_i=(p_i^x,p_i^y,p_i^z)$. The dynamic resource state is the residual-capacity vector $(f_1,\ldots,f_N)$; rack coordinates and pairwise distances are static. Chips within a rack are managed by a local allocator and connected by a high-speed local fabric. The global scheduler therefore chooses the participating racks and their chip allocations, while abstracting away chip-level placement within each rack.

\subsection{Job and Placement Model}\label{sec:job_model}
Each job $j$ is described by $(d_j,a_j,t_j)$: an integer chip demand $d_j>0$, arrival time $a_j$, and execution time $t_j$. Jobs are gang-scheduled, so a job receives all $d_j$ chips before starting and retains them until completion. A placement consists of a participating rack set $S_j\subseteq\mathcal{I}$ and allocations $x_j=\{x_{j,i}\}_{i\in S_j}$, where $x_{j,i}$ is the number of chips allocated on rack $i$. Feasibility requires

\begin{align}
x_{j,i} &\in \mathbb{Z}_{>0}, \quad x_{j,i}\leq f_i, \quad \forall i\in S_j, \label{eq:feasible_capacity}\\
\sum_{i\in S_j}x_{j,i} &= d_j. \label{eq:feasible_demand}
\end{align}
A job is rack-local feasible if $d_j\leq\max_i f_i$. If arbitrary residual capacities may be stitched, it is globally feasible if $d_j\leq\sum_i f_i$. The interval
\begin{equation}
\max_i f_i < d_j \leq \sum_i f_i
\label{eq:fragmentation_region}
\end{equation}
is the fragmentation regime of interest: no rack suffices alone, but the cluster has enough capacity in aggregate. A job with no policy-feasible placement at the current epoch is deferred in the waiting queue.

\subsection{OCS Ring-Cost Model}\label{sec:ocs_model}
For a cross-rack placement, the OCS fabric realizes $S_j=\{r_1,\ldots,r_k\}$ as a closed logical ring. The torus distance between racks $u$ and $v$ is
\begin{equation}
c(u,v)=\sum_{q\in\{x,y,z\}} \min\left(|p_u^q-p_v^q|,\ M_q-|p_u^q-p_v^q|\right).
\label{eq:torus_distance}
\end{equation}
For an ordering $\pi\in\Pi(S_j)$, with $\pi_{k+1}=\pi_1$, its realized ring cost and the ideal minimum are
\begin{align}
R(S_j,\pi) &= \sum_{m=1}^{k} c(\pi_m,\pi_{m+1}), \label{eq:ring_order_cost}\\
R^{\star}(S_j) &= \min_{\pi\in\Pi(S_j)} R(S_j,\pi), \label{eq:ring_cost}
\end{align}
where $\Pi(S_j)$ is the set of rack permutations. The closed-tour definition counts both directions when $k=2$. Ring cost does not model link contention or alter job execution time.

\subsection{Online Scheduling Semantics}
Scheduling is online and event-driven, with no knowledge of future arrivals. Execution time $t_j$ determines the completion event in the simulator but is not exposed to placement decisions. At each arrival or completion event, queued jobs are considered in arrival order. An infeasible job remains queued, but the scan continues, allowing later feasible jobs to start without head-of-line blocking. Once admitted, a job is non-preemptive and is neither migrated nor resized; it releases all allocated chips after $t_j$. We treat OCS configuration as an instantaneous logical operation and do not model port constraints, configuration failures, or concurrent-link contention. These abstractions isolate the effects of rack-level placement and topology-aware stitching.

\section{Design}
\begin{figure*}[t]
    \centering
    \includegraphics[width=1\textwidth]{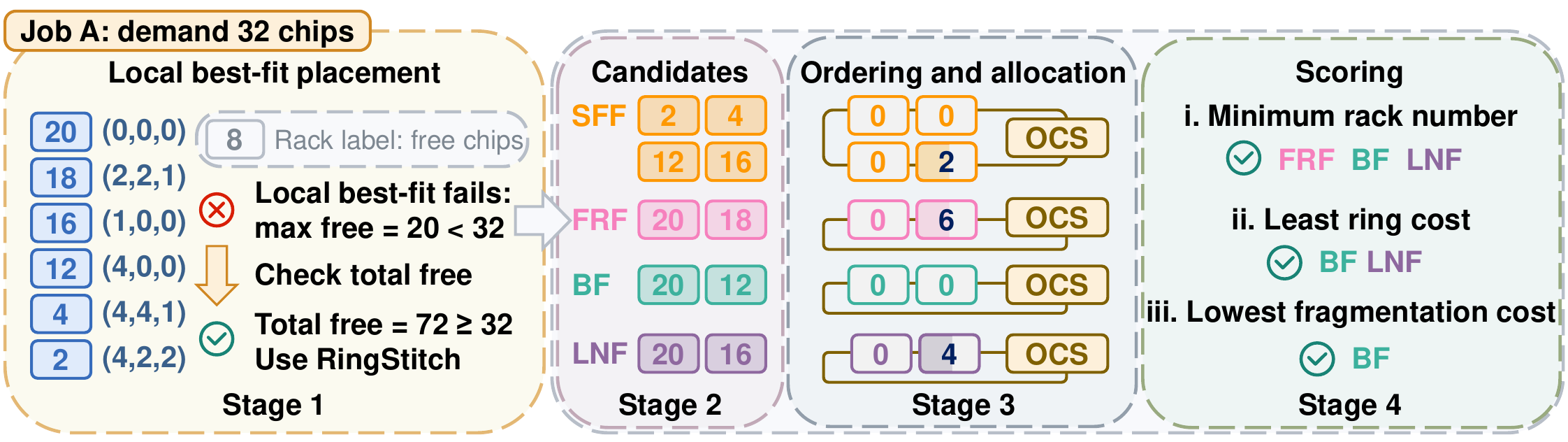}
    \caption{RingStitch placement workflow. Four heuristics generate at most four unique candidates: SFF (small-fragment-first), FRF (fewest-racks-first), BF (best-fit), and LNF (local-neighbor-first).}
    \label{fig:ringstitch_workflow}
\end{figure*}

\label{sec:ringstitch_design}
RingStitch is a local-first placement policy that recovers aggregate residual capacity while controlling rack count, ring cost, and post-placement fragmentation.

\subsection{Overview}
For each arriving job, RingStitch first attempts rack-local best-fit placement. Among racks with $f_i\geq d_j$, it selects the smallest $f_i$, minimizing the local leftover without invoking OCS. If no rack is locally feasible and $\sum_i f_i<d_j$, the job is deferred. Otherwise, RingStitch generates a bounded set of cross-rack candidates. For each candidate, it constructs a fragment-first chip allocation, computes a low-cost ring order, and evaluates a lexicographic score. It commits the minimum-score candidate and logically configures its saved ring order. Fig.~\ref{fig:ringstitch_workflow} illustrates this workflow, and Algorithm~\ref{alg:ringstitch} gives the per-job procedure. 
%Thus, cross-rack stitching is invoked only when local placement fails but aggregate residual capacity is sufficient.

\begin{algorithm}[tb]
    \small
    \caption{RingStitch placement for job $j$}
    \label{alg:ringstitch}
    \KwData{$\{f_i\}_{i=1}^{N}$, $c(u,v)$, and $d_j$}
    \KwResult{Placement $(S_j,x_j)$ with optional ring order $\pi_j$, or deferred}

    \If{$\exists i$ such that $f_i\ge d_j$}{
        select $i^\star = \arg\min_{i: f_i \ge d_j} f_i$\;
        place job $j$ in rack $i^\star$ without OCS stitching\;
        \Return local placement\;
    }
    \If{$\sum_i f_i < d_j$}{
        \Return deferred\;
    }
    $\mathcal{C}_j \leftarrow$ GenerateCandidates$(\{f_i\},d_j)$\;
    \ForEach{$S\in\mathcal{C}_j$}{
        build fragment-first allocation $x_j(S)$\;
        $(\widehat{\pi}(S),\widehat{R}(S))\leftarrow$ OrderRing$(S,c)$\;
        compute key $\left(|S|,\widehat{R}(S),F(S,x_j(S))\right)$\;
    }
    select $(S_j^{\star},x_j^{\star})$ with minimum lexicographic key\;
    logically configure OCS using $\widehat{\pi}(S_j^{\star})$\;
    commit allocation\;
    \Return $(S_j^{\star},x_j^{\star},\widehat{\pi}(S_j^{\star}))$\;
\end{algorithm}

\subsection{Hybrid Candidate Generation}
Enumerating all rack subsets is infeasible at SuperPod scale. RingStitch applies four complementary greedy heuristics to racks with positive residual capacity. Each heuristic stops once its selected capacity reaches $d_j$; infeasible and duplicate sets are removed, leaving at most four candidates in $\mathcal{C}_j$.

\noindent\textbf{Small-fragment-first (SFF).}
SFF scans racks in increasing order of $f_i$ and selects them until their aggregate capacity meets the demand. It exposes a candidate that consumes small residual blocks before larger ones.

\noindent\textbf{Fewest-racks-first (FRF).}
FRF scans racks in decreasing order of $f_i$. Taking the largest residual capacities first minimizes the number of racks needed to meet the demand and therefore bounds the number of ring participants.

\noindent\textbf{Best-fit (BF).}
Given the remaining demand, BF repeatedly chooses the largest residual capacity that does not exceed it; if every remaining rack is larger, it chooses the smallest such rack. This greedy rule seeks a close capacity fit but does not solve a global subset-sum problem.

\noindent\textbf{Local-neighbor-first (LNF).}
LNF starts from the rack with the largest $f_i$ and repeatedly adds the unselected rack whose minimum torus distance to the selected set is smallest. 

Together, the four heuristics expose cleanup, compactness, capacity-fit, and topology-local candidates without exhaustive subset enumeration.

\subsection{Allocation, Ring Ordering, and Scoring}
For every candidate, RingStitch first determines its chip allocation and ring realization, then scores the resulting placement.

\noindent\textbf{Fragment-first allocation.}
For a candidate set $S$, racks are processed in increasing order of residual capacity. If $P_i$ is the set processed before rack $i$, RingStitch assigns
\begin{equation}
x_{j,i}=\min\left(f_i,\ d_j-\sum_{h\in P_i}x_{j,h}\right),
\label{eq:fragment_alloc}
\end{equation}
and stops when the accumulated allocation reaches $d_j$. This rule exhausts small residual blocks when possible and preserves larger rack-level residual capacities for future jobs.

\noindent\textbf{Ring ordering.}
RingStitch computes a ring order $\widehat{\pi}(S)$ for every candidate before candidate selection. For $|S|\leq 8$, it fixes one rack, enumerates the remaining permutations, and obtains the exact minimum in Eq.~\eqref{eq:ring_cost}. For larger sets, it starts from the rack with the largest current residual capacity and repeatedly appends the rack nearest to the current ring tail, then closes the ring. This nearest-neighbor procedure produces a feasible, not necessarily optimal, order. Its realized cost is
\begin{equation}
\widehat{R}(S)=R(S,\widehat{\pi}(S)).
\label{eq:realized_ring_cost}
\end{equation}
Thus, $\widehat{R}(S)=R^{\star}(S)$ for small candidates, while $\widehat{R}(S)$ is a heuristic cost for larger ones.

\begin{figure*}[t]
\centering
\includegraphics[width=1\textwidth]
{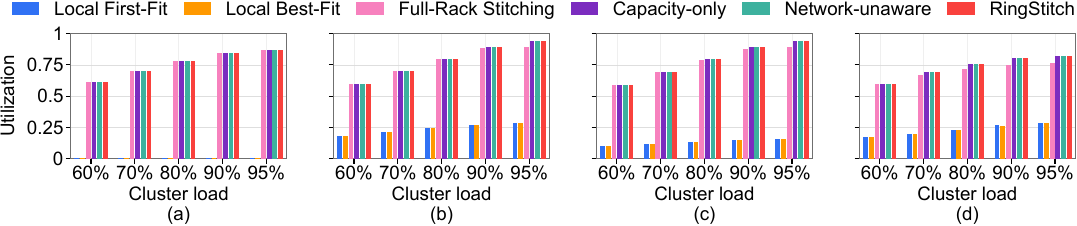}
\caption{Cluster utilization under Workloads A--D.}
\label{fig:utilization}
\end{figure*}

\noindent\textbf{Lexicographic scoring.}
Let $z_i=f_i-x_{j,i}$ be the residual capacity after the candidate allocation. RingStitch measures undesirable small leftovers with
\begin{align}
F(S,x_j) &= \sum_{i\in S}\phi(z_i), \label{eq:frag_cost}\\
\phi(z) &=
\begin{cases}
0, & z=0,\\
\log(1+1/z), & z>0.
\end{cases}
\end{align}
The penalty is largest for small positive leftovers, while a fully consumed residual block incurs zero cost. RingStitch selects
\begin{equation}
(S_j^{\star},x_j^{\star})=
\operatorname*{lex\,argmin}_{S\in\mathcal{C}_j}
\left(|S|,\ \widehat{R}(S),\ F(S,x_j(S))\right).
\label{eq:select_candidate}
\end{equation}
The ordering is strict: rack count is primary, realized ring cost breaks equal-cardinality choices, and fragmentation cost is the final tie-breaker. RingStitch therefore never adds a rack solely to lower ring or fragmentation cost. Exact ties follow candidate-generation order. After selection, it commits $x_j^{\star}$ and configures the previously computed $\widehat{\pi}(S_j^{\star})$.

\section{Evaluation}

We evaluate RingStitch with a discrete-event simulator that tracks rack-level residual capacity, job arrivals and completions, and OCS-aware placement decisions. We ask whether RingStitch (i) improves utilization, (ii) reduces blocking and waiting, and (iii) limits cross-rack communication cost.

\subsection{Methodology}
The simulated cluster follows the model in Sec.~\ref{sec:cluster_model}: $150$ racks, $64$ chips per rack, and $9600$ chips in total. Racks are embedded in a $5 \times 5 \times 6$ 3D torus, which is used to compute ring communication costs. Job arrivals follow a Poisson process, and job duration follows a size-dependent log-normal distribution. We vary the offered cluster load from $60\%$ to $95\%$ to cover both lightly loaded and fragmentation-heavy regimes. 

\begin{figure}[t]
\centering
\includegraphics[width=0.9\columnwidth]{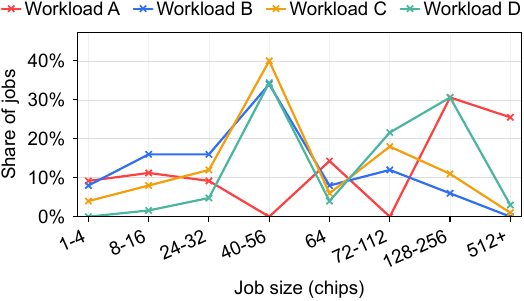}
\caption{Job-size distributions used in the evaluation.}
\label{fig:workloads}
\end{figure}

We use four workload mixes, shown in Fig.~\ref{fig:workloads}. Workload A is production-inspired and includes a sizable fraction of large slices. Workload B is a fragmented multi-tenant mix, dominated by 40--56-chip jobs with a smaller cross-rack tail. Workload C further increases 40--56-chip jobs and just-over-one-rack demands to stress fragmentation. Workload D is phased: long 40--56-chip jobs first create residual fragments, followed by many cross-rack jobs that amplify stitching pressure. Unless otherwise stated, each run uses a $50+200+50$ hour timeline: a $50$-hour warm-up phase, a $200$-hour measurement window, and a $50$-hour drain phase.

We repeat each configuration with 50 base random seeds. For each workload and seed, a common arrival stream is thinned to generate the load levels, and all schedulers see the same trace within each workload/load condition.

\subsection{Baselines}
We compare RingStitch with five baselines. \textit{Local First-Fit} scans racks in order and places a job in the first rack with enough residual capacity. \textit{Local Best-Fit} also restricts placement to one rack, but chooses the rack with the smallest leftover capacity. \textit{Full-Rack Stitching} models coarse-grained OCS scheduling: only completely free racks can participate in a stitched slice. \textit{Capacity-only Stitching} aggregates partial rack capacity until the demand is met, without considering topology or fragmentation. \textit{Network-unaware Best-Fit} greedily selects racks by capacity fit, ignoring communication cost. After selecting a cross-rack set, all cross-rack schedulers use the same ring-ordering procedure: exact enumeration for up to eight racks and a nearest-neighbor heuristic otherwise.

\subsection{Metrics}
We report four outcomes. Cluster utilization is the time-average fraction of busy chips during the measurement window. Blocking ratio is the chip-hour-weighted share of arriving demand not admitted by the window's end. Average and P95 waiting time include all measurement-window arrivals, with jobs not started by the waiting horizon assigned a capped wait from arrival to that horizon. Communication overhead is the mean 3D-torus ring cost among admitted cross-rack jobs; rack-local placements are excluded because they construct no cross-rack ring.

\subsection{Evaluation Results}

\noindent\textbf{Utilization.}
Fig.~\ref{fig:utilization} compares cluster utilization across workloads and load levels. Fragment-aware policies track the offered load much more closely than local and full-rack placement. At $95\%$ load, RingStitch achieves $94.5\%$ and $94.1\%$ utilization under Workloads B and C, exceeding Full-Rack Stitching by $5.0$ and $4.4$ percentage points; under Workload D, it reaches $82.1\%$, compared with $76.9\%$ for Full-Rack Stitching and $28.6\%$ for Local First-Fit. RingStitch matches Capacity-only Stitching and Network-unaware Best-Fit at every point. Local placement is confined to one rack, while Full-Rack Stitching cannot combine partial capacity; RingStitch recovers these fragments without losing the utilization of aggregate-capacity stitching.

\begin{figure*}[t]
\centering
\includegraphics[width=1\textwidth]{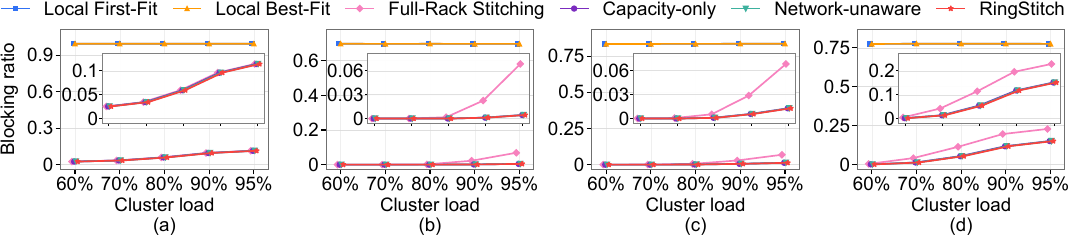}
\caption{Demand-weighted blocking ratio under Workloads A--D.}
\label{fig:blocking}
\end{figure*}

\noindent\textbf{Blocking.}
Fig.~\ref{fig:blocking} compares demand-weighted blocking under the same settings. At $95\%$ load, RingStitch lowers blocking from $6.9\%$ to $0.4\%$ in Workload B, from $6.8\%$ to $1.3\%$ in Workload C, and from $22.9\%$ to $15.1\%$ in Workload D relative to Full-Rack Stitching; the corresponding local-only ratios are $69.7\%$, $83.5\%$, and $77.5\%$. RingStitch matches the other fragment-aware policies because these policies share aggregate-capacity feasibility. The observed gap to Full-Rack Stitching isolates allocation-granularity blocking: long 40--56-chip jobs leave partial capacity that whole-rack stitching cannot use, especially when Workload D's later cross-rack jobs encounter fragments retained by its long-running early phase. RingStitch reclaims these fragments but cannot eliminate genuine capacity shortfalls.

\begin{table}[t]
\centering
\caption{Average and P95 waiting time under Workloads A--D. Each cell reports Avg/P95 waiting time in hours.}
\label{tab:waiting_time}
\scriptsize
\setlength{\tabcolsep}{1.8pt}
\renewcommand{\arraystretch}{1.05}
\definecolor{ringstitchlavender}{RGB}{232,229,249}
\begin{tabular}{lccccc>{\columncolor{ringstitchlavender}}c}
\toprule
Workload & Local FF & Local BF & Full-Rack & Capacity-only & Net-unaware & RingStitch \\
\midrule
\multirow{5}{*}{A}\quad 0.60 & 85.9/232.0 & 85.9/232.0 &  0.6/4.0  &  0.6/4.0  &  0.6/4.0  &  0.6/4.0  \\
\phantom{A}\quad 0.70 & 86.0/232.6 & 86.0/232.6 &  1.0/6.1  &  1.0/6.2  &  1.0/6.2  &  1.0/6.2  \\
\phantom{A}\quad 0.80 & 85.4/232.4 & 85.4/232.4 &  1.7/10.3 &  1.7/10.3 &  1.7/10.3 &  1.7/10.3 \\
\phantom{A}\quad 0.90 & 84.8/232.4 & 84.8/232.4 &  2.6/16.5 &  2.6/16.5 &  2.6/16.5 &  2.6/16.5 \\
\phantom{A}\quad 0.95 & 84.9/232.3 & 84.9/232.3 &  3.2/19.5 &  3.2/19.5 &  3.2/19.5 &  3.2/19.5 \\
\midrule
\multirow{5}{*}{B}\quad 0.60 & 27.0/194.4 & 27.0/194.4 &  0.0/0.0  &  0.0/0.0  &  0.0/0.0  &  0.0/0.0  \\
\phantom{B}\quad 0.70 & 27.0/194.6 & 27.0/194.6 &  0.0/0.0  &  0.0/0.0  &  0.0/0.0  &  0.0/0.0  \\
\phantom{B}\quad 0.80 & 27.0/194.5 & 27.0/194.5 &  0.1/0.3  &  0.0/0.0  &  0.0/0.0  &  0.0/0.0  \\
\phantom{B}\quad 0.90 & 27.0/194.3 & 27.0/194.3 &  0.8/5.3  &  0.1/0.3  &  0.1/0.3  &  0.1/0.3  \\
\phantom{B}\quad 0.95 & 27.0/194.3 & 27.0/194.3 &  2.2/15.4 &  0.2/0.8  &  0.2/0.8  &  0.2/0.8  \\
\midrule
\multirow{5}{*}{C}\quad 0.60 & 44.8/215.7 & 44.8/215.7 &  0.0/0.0  &  0.0/0.0  &  0.0/0.0  &  0.0/0.0  \\
\phantom{C}\quad 0.70 & 44.7/215.5 & 44.7/215.5 &  0.0/0.1  &  0.0/0.0  &  0.0/0.0  &  0.0/0.0  \\
\phantom{C}\quad 0.80 & 44.7/215.6 & 44.7/215.6 &  0.1/0.7  &  0.0/0.1  &  0.0/0.1  &  0.0/0.1  \\
\phantom{C}\quad 0.90 & 44.8/215.8 & 44.8/215.8 &  0.7/4.1  &  0.1/0.7  &  0.1/0.7  &  0.1/0.7  \\
\phantom{C}\quad 0.95 & 44.9/216.1 & 44.9/216.1 &  1.7/8.4  &  0.4/1.7  &  0.4/1.7  &  0.4/1.7  \\
\midrule
\multirow{5}{*}{D}\quad 0.60 & 60.2/158.6 & 60.2/158.6 &  0.2/0.9  &  0.0/0.2  &  0.0/0.2  &  0.0/0.2  \\
\phantom{D}\quad 0.70 & 60.4/158.8 & 60.4/158.8 &  1.0/4.8  &  0.3/1.9  &  0.3/1.9  &  0.3/1.9  \\
\phantom{D}\quad 0.80 & 60.4/158.8 & 60.4/158.8 &  2.9/10.8 &  1.2/6.0  &  1.2/6.0  &  1.2/6.0  \\
\phantom{D}\quad 0.90 & 60.7/158.7 & 60.6/158.7 &  5.0/15.8 &  2.5/11.1 &  2.5/11.1 &  2.5/11.1 \\
\phantom{D}\quad 0.95 & 60.8/158.7 & 60.8/158.7 &  6.4/18.5 &  3.2/13.8 &  3.2/13.8 &  3.2/13.8 \\
\bottomrule
\end{tabular}
\end{table}

\noindent\textbf{Waiting time.}
Table~\ref{tab:waiting_time} compares average and P95 waiting time. At $95\%$ load in Workload B, RingStitch records $0.2/0.8$ hours, compared with $2.2/15.4$ for Full-Rack Stitching and $27.0/194.3$ for either local scheduler; under Workload D, the corresponding values are $3.2/13.8$, $6.4/18.5$, and about $60.8/158.7$ hours. In Workload B, these results represent $91.6\%$ and $94.7\%$ reductions relative to Full-Rack Stitching. Jobs exceeding one rack are infeasible for local policies and accumulate large capped waits, while Full-Rack Stitching must await entire free racks. RingStitch instead starts them from partial capacity; its equality with Capacity-only Stitching and Network-unaware Best-Fit shows that topology control adds no waiting-time penalty.

\begin{figure*}[t]
\centering
\includegraphics[width=1\textwidth]{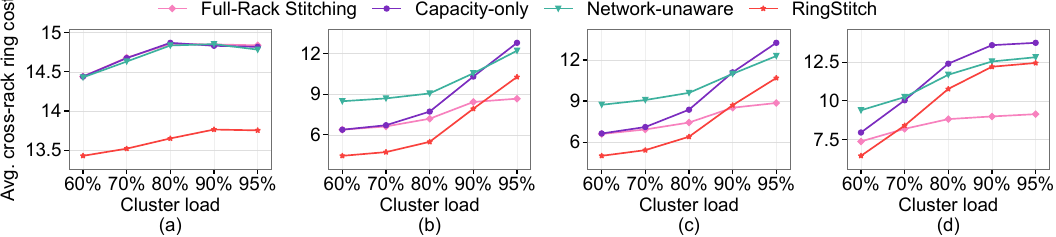}
\caption{Average ring communication cost among admitted cross-rack jobs under Workloads A--D.}
\label{fig:ring_cost}
\end{figure*}

\noindent\textbf{Communication cost.}
Fig.~\ref{fig:ring_cost} compares the four cross-rack schedulers; local policies are omitted because they never construct a ring. At $95\%$ load in Workload B, RingStitch costs $10.26$, compared with $12.78$ for Capacity-only Stitching and $12.20$ for Network-unaware Best-Fit, reductions of $19.7\%$ and $15.9\%$. Capacity-only favors large residuals, and Network-unaware favors capacity fit, so both may select dispersed racks. RingStitch uses topology-local candidates and favors compact, low-cost rack sets, retaining the same admission behavior with less communication. Full-Rack Stitching can have lower conditional cost in Workloads B--D because it uses completely free racks and admits a different job set; this result must be read together with its lower utilization, higher blocking ratio and longer waiting time.

\subsection{Takeaways}
RingStitch turns partial rack capacity into schedulable resources, overcoming the placement granularity that limits local and full-rack policies. Across all workloads and load levels, it matches Capacity-only Stitching and Network-unaware Best-Fit in utilization, blocking, and waiting time while consistently reducing cross-rack ring cost. These results show that topology-aware fragment stitching can recover otherwise stranded capacity with lower modeled ring cost than topology-unaware aggregation.

\section{Related Work}

\textbf{Optically reconfigurable AI infrastructure.}
TPU v4 shows that optical circuit switches can construct flexible training topologies by selecting hardware units and configuring them into logical 3D tori, improving placement flexibility, availability, and utilization~\cite{jouppi2023tpu}. Recent TPU generations further scale this model, making optical reconfiguration increasingly important for resource organization in large AI clusters~\cite{tpu8t}. More broadly, Matryoshka revisits hyperscale datacenter design, while OpenOptics provides open implementation support for optical datacenter networks~\cite{cai2026matryoshka,lei2026openoptics}. RingStitch shares this optical-AI infrastructure setting but focuses on recovering fragmented rack capacity rather than allocating only complete racks or hardware blocks.

\textbf{Reconfigurable datacenter networks.}
Optical and hybrid datacenter networks, including c-Through, Helios, OSA, RotorNet, Opera, and Sirius, use optical circuits or fast topology changes to improve bandwidth efficiency and topology flexibility~\cite{wang2010c,farrington2010helios,chen2013osa,mellette2017rotornet,mellette2020expanding,ballani2020sirius}. Recent routing work further explores how to exploit path diversity and topology dynamics in reconfigurable optical datacenter networks~\cite{li2024uniform,li2025unlocking}. These systems mainly optimize network architecture, routing, or traffic scheduling, whereas RingStitch uses optical reconfiguration as a placement-time defragmentation mechanism.

\textbf{Deep learning cluster scheduling.}
GPU-cluster schedulers such as Gandiva, Tiresias, Themis, Pollux, Gavel, Optimus, and HiveD improve job completion time, fairness, goodput, heterogeneity support, or sharing guarantees~\cite{xiao2018gandiva,gu2019tiresias,mahajan2020themis,qiao2021pollux,narayanan2020heterogeneity,peng2018optimus,zhao2020hived}. CASSINI shows that network-aware placement can improve ML-cluster efficiency~\cite{rajasekaran2024cassini}, while RailS optimizes load balancing for all-to-all communication in distributed MoE training~\cite{xu2026rails}. These works optimize scheduling or communication under conventional cluster abstractions; they do not model OCS as a resource for stitching partial rack fragments into communication-aware slices.

%In contrast, RingStitch treats optical reconfiguration as a demand-aware defragmentation mechanism for partial rack capacity in multi-tenant TPU-style clusters.

\section{Conclusion}
This paper presented RingStitch, a demand-aware optical defragmentation scheduler for multi-tenant TPU-style clusters. RingStitch identifies fragmented rack capacity, selects compact fragment-aware rack sets, and stitches them into low-cost logical OCS rings. By elevating partial rack capacity to a schedulable resource, RingStitch reduces fragmentation-induced blocking while achieving lower modeled ring cost than naive cross-rack placement. The design opens a practical direction for using optical reconfiguration not only for large training slices but also for fine-grained defragmentation in multi-tenant AI clusters.

\bibliographystyle{unsrt}
\bibliography{reference}

%\vfill

\end{document}